\documentclass[preprint]{revtex4}
\usepackage{graphicx}

\newcommand*{\pt}{\ensuremath{p_{\mathrm{T}}}}
\newcommand*{\mub}{\ensuremath{\mu_{B}}}
\newcommand*{\dedx}{\ensuremath{\mathrm{d}E/\mathrm{d}x}}
\newcommand*{\ktopi}{\ensuremath{\mathrm{K}/\pi}}
\newcommand*{\ptopi}{\ensuremath{(\mathrm{p} + \bar{\mathrm{p}})/\pi}}
\newcommand*{\ktop}{\ensuremath{(\mathrm{K}^{+}\!+\mathrm{K}^{-})/(\mathrm{p}+\bar{\mathrm{p}})}}
\newcommand*{\ktopplus}{\ensuremath{\mathrm{K}^{+}\!/\mathrm{p}}}
\newcommand*{\roots}{\ensuremath{\sqrt{s_{_{NN}}}}}
\newcommand*{\sdyn}{\ensuremath{\sigma_{\mathrm{dyn}}}}
\newcommand*{\sdata}{\ensuremath{\sigma_{\mathrm{data}}}}
\newcommand*{\smix}{\ensuremath{\sigma_{\mathrm{mix}}}}

\newcommand*{\cbs}{\ensuremath{C_{BS}}}

\begin{document}

\title{Energy dependence of kaon-to-proton ratio fluctuations in central Pb+Pb collisions from \roots\ = 6.3 to 17.3~GeV}


\affiliation{NIKHEF, Amsterdam, Netherlands.}
\affiliation{Department of Physics, University of Athens, Athens, Greece.}
\affiliation{E\"otv\"os Lor\'ant University, Budapest, Hungary.}
\affiliation{KFKI Research Institute for Particle and Nuclear Physics, Budapest, Hungary.}
\affiliation{MIT, Cambridge, USA.}
\affiliation{H.~Niewodnicza\'nski Institute of Nuclear Physics, Polish Academy of Sciences, Cracow, Poland.}
\affiliation{Gesellschaft f\"{u}r Schwerionenforschung (GSI), Darmstadt, Germany.}
\affiliation{Joint Institute for Nuclear Research, Dubna, Russia.}
\affiliation{Fachbereich Physik der Universit\"{a}t, Frankfurt, Germany.}
\affiliation{CERN, Geneva, Switzerland.}
\affiliation{Institute of Physics, Jan Kochanowski University, Kielce, Poland.}
\affiliation{Fachbereich Physik der Universit\"{a}t, Marburg, Germany.}
\affiliation{Max-Planck-Institut f\"{u}r Physik, Munich, Germany.}
\affiliation{Inst. of Particle and Nuclear Physics, Charles Univ., Prague, Czech Republic.}
\affiliation{Nuclear Physics Laboratory, University of Washington, Seattle, WA, USA.}
\affiliation{Atomic Physics Department, Sofia Univ. St. Kliment Ohridski, Sofia, Bulgaria.}
\affiliation{Institute for Nuclear Research and Nuclear Energy, BAS, Sofia, Bulgaria.}
\affiliation{Department of Chemistry, Stony Brook Univ. (SUNYSB), Stony Brook, USA.}
\affiliation{Institute for Nuclear Studies, Warsaw, Poland.}
\affiliation{Institute for Experimental Physics, University of Warsaw, Warsaw, Poland.}
\affiliation{Faculty of Physics, Warsaw University of Technology, Warsaw, Poland.}
\affiliation{Rudjer Boskovic Institute, Zagreb, Croatia.}

\author{T.~Anticic}
\affiliation{Rudjer Boskovic Institute, Zagreb, Croatia.}
\author{B.~Baatar}
\affiliation{Joint Institute for Nuclear Research, Dubna, Russia.}
\author{D.~Barna}
\affiliation{KFKI Research Institute for Particle and Nuclear Physics, Budapest, Hungary.}
\author{J.~Bartke}
\affiliation{H.~Niewodnicza\'nski Institute of Nuclear Physics, Polish Academy of Sciences, Cracow, Poland.}
\author{H.~Beck}
\affiliation{Fachbereich Physik der Universit\"{a}t, Frankfurt, Germany.}
\author{L.~Betev}
\affiliation{CERN, Geneva, Switzerland.}
\author{H.~Bia{\l}\-kowska}
\affiliation{Institute for Nuclear Studies, Warsaw, Poland.}
\author{C.~Blume}
\affiliation{Fachbereich Physik der Universit\"{a}t, Frankfurt, Germany.}
\author{M.~Bogusz}
\affiliation{Faculty of Physics, Warsaw University of Technology, Warsaw, Poland.}
\author{B.~Boimska}
\affiliation{Institute for Nuclear Studies, Warsaw, Poland.}
\author{J.~Book}
\affiliation{Fachbereich Physik der Universit\"{a}t, Frankfurt, Germany.}
\author{M.~Botje}
\affiliation{NIKHEF, Amsterdam, Netherlands.}
\author{P.~Bun\v{c}i\'{c}}
\affiliation{CERN, Geneva, Switzerland.}
\author{T.~Cetner}
\affiliation{Faculty of Physics, Warsaw University of Technology, Warsaw, Poland.}
\author{P.~Christakoglou}
\affiliation{NIKHEF, Amsterdam, Netherlands.}
\author{P.~Chung}
\affiliation{Department of Chemistry, Stony Brook Univ. (SUNYSB), Stony Brook, USA.}
\author{O.~Chvala}
\affiliation{Inst. of Particle and Nuclear Physics, Charles Univ., Prague, Czech Republic.}
\author{J.G.~Cramer}
\affiliation{Nuclear Physics Laboratory, University of Washington, Seattle, WA, USA.}
\author{V.~Eckardt}
\affiliation{Max-Planck-Institut f\"{u}r Physik, Munich, Germany.}
\author{Z.~Fodor}
\affiliation{KFKI Research Institute for Particle and Nuclear Physics, Budapest, Hungary.}
\author{P.~Foka}
\affiliation{Gesellschaft f\"{u}r Schwerionenforschung (GSI), Darmstadt, Germany.}
\author{V.~Friese}
\affiliation{Gesellschaft f\"{u}r Schwerionenforschung (GSI), Darmstadt, Germany.}
\author{M.~Ga\'zdzicki}
\affiliation{Fachbereich Physik der Universit\"{a}t, Frankfurt, Germany.}
\affiliation{Institute of Physics, Jan Kochanowski University, Kielce, Poland.}
\author{K.~Grebieszkow}
\affiliation{Faculty of Physics, Warsaw University of Technology, Warsaw, Poland.}
\author{C.~H\"{o}hne}
\affiliation{Gesellschaft f\"{u}r Schwerionenforschung (GSI), Darmstadt, Germany.}
\author{K.~Kadija}
\affiliation{Rudjer Boskovic Institute, Zagreb, Croatia.}
\author{A.~Karev}
\affiliation{CERN, Geneva, Switzerland.}
\author{V.I.~Kolesnikov}
\affiliation{Joint Institute for Nuclear Research, Dubna, Russia.}
\author{T.~Kollegger}
\affiliation{Fachbereich Physik der Universit\"{a}t, Frankfurt, Germany.}
\author{M.~Kowalski}
\affiliation{H.~Niewodnicza\'nski Institute of Nuclear Physics, Polish Academy of Sciences, Cracow, Poland.}
\author{D.~Kresan}
\affiliation{Gesellschaft f\"{u}r Schwerionenforschung (GSI), Darmstadt, Germany.}
\author{A.~Laszlo}
\affiliation{KFKI Research Institute for Particle and Nuclear Physics, Budapest, Hungary.}
\author{R.~Lacey}
\affiliation{Department of Chemistry, Stony Brook Univ. (SUNYSB), Stony Brook, USA.}
\author{M.~van~Leeuwen}
\affiliation{NIKHEF, Amsterdam, Netherlands.}
\author{M.~Mackowiak}
\affiliation{Faculty of Physics, Warsaw University of Technology, Warsaw, Poland.}
\author{M.~Makariev}
\affiliation{Institute for Nuclear Research and Nuclear Energy, BAS, Sofia, Bulgaria.}
\author{A.I.~Malakhov}
\affiliation{Joint Institute for Nuclear Research, Dubna, Russia.}
\author{M.~Mateev}
\affiliation{Atomic Physics Department, Sofia Univ. St. Kliment Ohridski, Sofia, Bulgaria.}
\author{G.L.~Melkumov}
\affiliation{Joint Institute for Nuclear Research, Dubna, Russia.}
\author{M.~Mitrovski}
\affiliation{Fachbereich Physik der Universit\"{a}t, Frankfurt, Germany.}
\author{St.~Mr\'owczy\'nski}
\affiliation{Institute of Physics, Jan Kochanowski University, Kielce, Poland.}
\author{V.~Nicolic}
\affiliation{Rudjer Boskovic Institute, Zagreb, Croatia.}
\author{G.~P\'{a}lla}
\affiliation{KFKI Research Institute for Particle and Nuclear Physics, Budapest, Hungary.}
\author{A.D.~Panagiotou}
\affiliation{Department of Physics, University of Athens, Athens, Greece.}
\author{W.~Peryt}
\affiliation{Faculty of Physics, Warsaw University of Technology, Warsaw, Poland.}
\author{J.~Pluta}
\affiliation{Faculty of Physics, Warsaw University of Technology, Warsaw, Poland.}
\author{D.~Prindle}
\affiliation{Nuclear Physics Laboratory, University of Washington, Seattle, WA, USA.}
\author{F.~P\"{u}hlhofer}
\affiliation{Fachbereich Physik der Universit\"{a}t, Marburg, Germany.}
\author{R.~Renfordt}
\affiliation{Fachbereich Physik der Universit\"{a}t, Frankfurt, Germany.}
\author{C.~Roland}
\affiliation{MIT, Cambridge, USA.}
\author{G.~Roland}
\affiliation{MIT, Cambridge, USA.}
\author{M. Rybczy\'nski}
\affiliation{Institute of Physics, Jan Kochanowski University, Kielce, Poland.}
\author{A.~Rybicki}
\affiliation{H.~Niewodnicza\'nski Institute of Nuclear Physics, Polish Academy of Sciences, Cracow, Poland.}
\author{A.~Sandoval}
\affiliation{Gesellschaft f\"{u}r Schwerionenforschung (GSI), Darmstadt, Germany.}
\author{N.~Schmitz}
\affiliation{Max-Planck-Institut f\"{u}r Physik, Munich, Germany.}
\author{T.~Schuster}
\affiliation{Fachbereich Physik der Universit\"{a}t, Frankfurt, Germany.}
\author{P.~Seyboth}
\affiliation{Max-Planck-Institut f\"{u}r Physik, Munich, Germany.}
\author{F.~Sikl\'{e}r}
\affiliation{KFKI Research Institute for Particle and Nuclear Physics, Budapest, Hungary.}
\author{E.~Skrzypczak}
\affiliation{Institute for Experimental Physics, University of Warsaw, Warsaw, Poland.}
\author{M.~Slodkowski}
\affiliation{Faculty of Physics, Warsaw University of Technology, Warsaw, Poland.}
\author{G.~Stefanek}
\affiliation{Institute of Physics, Jan Kochanowski University, Kielce, Poland.}
\author{R.~Stock}
\affiliation{Fachbereich Physik der Universit\"{a}t, Frankfurt, Germany.}
\author{H.~Str\"{o}bele}
\affiliation{Fachbereich Physik der Universit\"{a}t, Frankfurt, Germany.}
\author{T.~Susa}
\affiliation{Rudjer Boskovic Institute, Zagreb, Croatia.}
\author{M.~Szuba}
\affiliation{Faculty of Physics, Warsaw University of Technology, Warsaw, Poland.}
\author{M.~Utvi\'{c}}
\affiliation{Fachbereich Physik der Universit\"{a}t, Frankfurt, Germany.}
\author{D.~Varga}
\affiliation{E\"otv\"os Lor\'ant University, Budapest, Hungary.}
\author{M.~Vassiliou}
\affiliation{Department of Physics, University of Athens, Athens, Greece.}
\author{G.I.~Veres}
\affiliation{KFKI Research Institute for Particle and Nuclear Physics, Budapest, Hungary.}
\author{G.~Vesztergombi}
\affiliation{KFKI Research Institute for Particle and Nuclear Physics, Budapest, Hungary.}
\author{D.~Vrani\'{c}}
\affiliation{Gesellschaft f\"{u}r Schwerionenforschung (GSI), Darmstadt, Germany.}
\author{Z.~W{\l}odarczyk}
\affiliation{Institute of Physics, Jan Kochanowski University, Kielce, Poland.}
\author{A.~Wojtaszek-Szwarc}
\affiliation{Institute of Physics, Jan Kochanowski University, Kielce, Poland.}

\collaboration{The NA49 collaboration}
\noaffiliation

\date{\today}


\begin{abstract}

Kaons and protons carry large parts of two conserved quantities, strangeness and baryon number. It is argued that their correlation and thus also fluctuations are sensitive to conditions prevailing at the anticipated parton-hadron phase boundary.
Fluctuations of the \ktop\ and \ktopplus\ ratios have been measured for the first time by NA49 in central Pb+Pb collisions at 5 SPS energies between \roots = 6.3~GeV and 17.3~GeV.
Both ratios exhibit a change of sign in \sdyn, a measure of non-statistical fluctuations, around \roots\ = 8~GeV.
Below this energy, \sdyn\ is positive, indicating higher fluctuation compared to a mixed event background sample, while for higher energies, \sdyn\ is negative, indicating correlated emission of kaons and protons.
The results are compared to hadronic transport model calculations which fail to reproduce the energy dependence.

\end{abstract}

\maketitle


Heavy-ion collisions serve as the laboratory to study hadronic matter under extreme energy density and temperature conditions. They surpass critical values of energy density where lattice QCD calculations predict a phase transition from hadronic to deconfined matter (the ``quark-gluon plasma'')~\cite{Karsch:2003jg}.
A multitude of relevant physics observables provides indication~\cite{Arnaldi:2008fw,Heinz:2000bk,Gazdzicki:1998vd,Afanasiev:2002mx,:2007fe,Friese:CPOD09,Gazdzicki:2004ef} that this phase transition might first occur in the energy range of the CERN SPS ($6.3~\mathrm{GeV} \leq \roots \leq 17.3~\mathrm{GeV}$): the domain of the present study.

While the above evidence is based on inclusive observables, additional insight can be obtained by looking at event-by-event fluctuations. Enhanced fluctuations are a general feature of phase transitions. Recent lattice QCD calculations report indications of a QCD critical point at finite baryo-chemical potential, reflected in a steep rise of quark number density fluctuation (quark ``susceptibility'')~\cite{Fodor:2004nz,Ejiri:2003dc,Karsch:2007dp}.
In addition, the changed correlation patterns in particle production, expected where new degrees of freedom emerge, are accessible via fluctuation observables~\cite{Koch:2008ia}.

Numerous such observables have been explored previously~\cite{Koch:2008ia,Zaranek:2001di,Alt:2004ir}, in part with inconclusive outcome. This resulted from uncontrolled sources of background fluctuations, notably variation of the reaction volume due to concurrent impact geometry fluctuations, finite number statistics, or from an obliteration of the signals during the hadron-resonance expansion phase that follows hadron formation at the parton-hadron phase boundary (notably so for electric charge correlations~\cite{Zaranek:2001di,Alt:2004ir}).

The present study of hadron ratio fluctuations overcomes the mentioned difficulties to a certain extent.
Effects from volume fluctuations are minimized, as a particle ratio is an intensive quantity and the average hadron ratios change only marginally in the analysed centrality interval.
Fluctuations from finite number statistics, as well as effects of the limited detector resolution in the particle identification (PID) method are removed by subtracting a mixed event reference. Thus, \emph{dynamical} fluctuations are studied.

For kaon-to-proton number ratio fluctuations, the conserved charges strangeness and baryon number are carried by heavy particles, making their dispersion in momentum space smaller compared to the case of electric charge fluctuations, where the $Q$-value of resonance decays creates a noticeable difference between the original charge distribution and that of the finally observed pions~\cite{Zaranek:2001di}. No such transport process (as, e.g.\ a resonance feeding into $\mathrm{K}^{+}\! + \mathrm{p}$) is known in the case discussed here.

The results presented here are the first attempt to experimentally probe the baryon number-strangeness correlation, which undergoes a massive change at the deconfinement phase transition.
Above that transition, where strange quarks ($S=-1$, $B=1/3$) are the relevant degrees of freedom, strangeness $S$ can only exist in direct conjunction with baryon number $B$. In a hadron gas, kaons ($S=-1$, $B=0$) allow for strangeness number unrelated to baryon number.
An appropriate correlation coefficient \cbs~\cite{Koch:2005vg} has been proposed which is related to the above mentioned quark number density susceptibililties at finite baryo-chemical potential \mub\ recently predicted by lattice QCD~\cite{Fodor:2004nz,Ejiri:2003dc,Karsch:2007dp}. The latter exhibit steep maxima at $T=150~\mathrm{MeV}$ and $\mub \approx 400~\mathrm{MeV}$, conditions that are reached in A+A collisions in the energy range under investigation in this analysis.
An important contribution to this coefficient are the kaon-proton correlations, that are experimentally probed for the first time in the present data analysis.
In a dense hadronic medium, in contrast to a gas of uncorrelated hadrons, a variety of processes may lead to a correlation of baryon number and strangeness, e.g.\ the associated production of hyperons and strange mesons. The effect of these hadronic correlations on \cbs\ was studied in the hadronic transport model UrQMD~\cite{Bleicher:1999xi,Bass:1998ca,Petersen:2008kb} and found to be small and constant in the SPS energy range.


The following analysis is based on data from Pb+Pb collisions recorded at the CERN SPS at five energies, $\roots = 6.3, 7.6, 8.7, 12.3$ and $17.3$~GeV. The $3.5\%$ most central events were selected, also avoiding an influence of the change in the inclusive $\langle \mathrm{K} \rangle / \langle \mathrm{p} \rangle$ ratio with centrality.

\begin{figure}
\includegraphics[width=8cm]{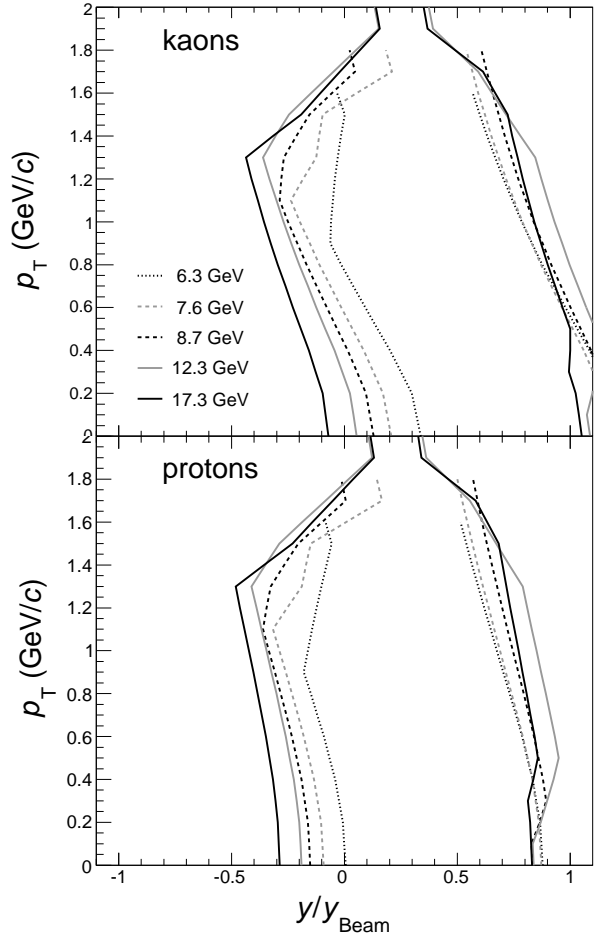}
\caption{
The acceptance for kaons and protons, as used in the present analysis, as a function of transverse momentum and the center-of-mass rapidity, normalized by the corresponding beam rapidity, for all analysed energies.
Lines delimit the regions in which particles can be identified.
Limits are due to geometric acceptance and available statistics, the latter predominantly at large momenta $p$ and transverse momenta~\pt.}
\label{fig:acc}
\end{figure}

The NA49 detector~\cite{Afanasev:1999iu} uses 4 large volume time projection chambers (TPC) for tracking and particle identification (PID) via their specific energy loss (\dedx) in the TPC gas.
During the SPS energy scan program, care was taken to keep the acceptance approximately constant with respect to midrapidity by setting the magnetic field strength proportional to the beam momentum.
To ensure best resolution and stability of the \dedx\ measurement, only tracks within the acceptance of the large main TPCs were accepted for the analysis. Further quality cuts are applied on the distance of closest approach of the extrapolated particle trajectory back to the main vertex and on the number of measured \dedx\ samples on the track.
The resulting acceptance as a function of center-of-mass rapidity normalized by beam rapidity, and transverse momentum \pt\ is depicted in Fig.~\ref{fig:acc}. The azimuthal acceptance is described in~\cite{:2008ca} and the comprehensive acceptance tables to be used in simulations can be found in~\cite{acc_tab_edms}. At each energy, the largest possible acceptance was used to ensure the highest statistical significance. Average uncorrected multiplicities within this acceptance are given in table~\ref{tbl:mult}.

\begin{table}
\begin{ruledtabular}
\begin{tabular}{*{5}{c}}
	\roots (GeV) & $\langle \mathrm{K}^+ \rangle$ & $\langle \mathrm{K}^- \rangle$ &  $\langle \mathrm{p} \rangle$ & $\langle \bar{\mathrm{p}} \rangle$ \\
	\hline
	6.3 & 5 & 1 & 28 & 0 \\
	7.6 & 8 & 2 & 35 & 0 \\
	8.7 & 12 & 4 & 41 & 0 \\
	12.3 & 22 & 9 & 53 & 1 \\
	17.3 & 34 & 20 & 72 & 3 \\
\end{tabular}
\end{ruledtabular}
\caption{Average uncorrected multiplicities within the acceptance of the present study.}
\label{tbl:mult}
\end{table}

The analysis procedure is similar to the one employed by NA49 in the study of fluctuations of the $\mathrm{K}/\pi$ and $\mathrm{p}/\pi$ ratios~\cite{:2008ca}.
The \dedx\ resolution of approximately 4\% allows for a statistical determination of the inclusive particle yields through a $\chi^2$ fit to the \dedx\ spectra in momentum space bins.
Probability density functions $f_m$ for \dedx\ and momentum $F_m$ from this inclusive analysis are then used as input for the event-wise particle ratio determination. The \dedx\ distributions overlap for different hadron species $m$, making a simple particle counting impossible. Therefore, an unbinned likelihood method as introduced in~\cite{Afanasev:2000fu,Gazdzicki:1994vj} is used.

In each event, the likelihood function $L$ is obtained by multiplying the probabilities of the $n$ particles in the event ($\langle n \rangle \approx 60$ at $\roots = 6.3$~GeV and $\langle n \rangle \approx 600$ at $\roots = 17.3$~GeV):
\begin{equation}
L = \prod_{i=1}^n \left[ \sum_m \Theta_m F_m \left( \vec{p_i} \right) f_m \left( \vec{p_i}, (\dedx)_i \right) \right],
\end{equation}
where $\Theta_m$ are the relative abundances, constructed such that $\sum_m \Theta_m = 1$. They are the only free parameters in the subsequent likelihood maximization. The event-wise hadron ratios are then calculated from the fitted values of $\Theta_m$.

\begin{figure}
\includegraphics[width=8cm]{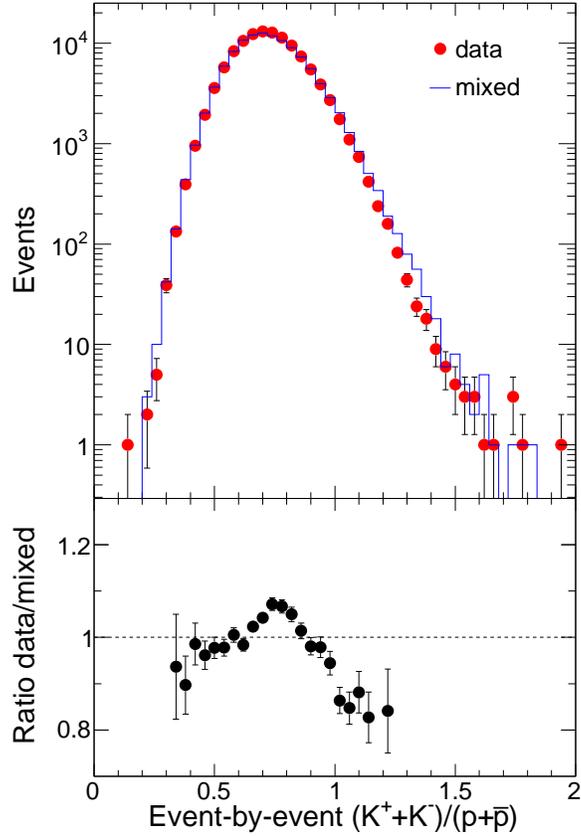}
\caption{(Color online) Event-by-event distribution of the \ktop\ ratio at \roots = 17.3~GeV for real data events compared to the mixed event reference. The lower panel shows the ratio data/mixed, where the convex shape indicates negative dynamical fluctuations. Only in the ratio plot, for better readability, statistically insignificant bins are not shown.}
\label{fig:ebye_ktop_160}
\end{figure}

\begin{figure}
\includegraphics[width=8cm]{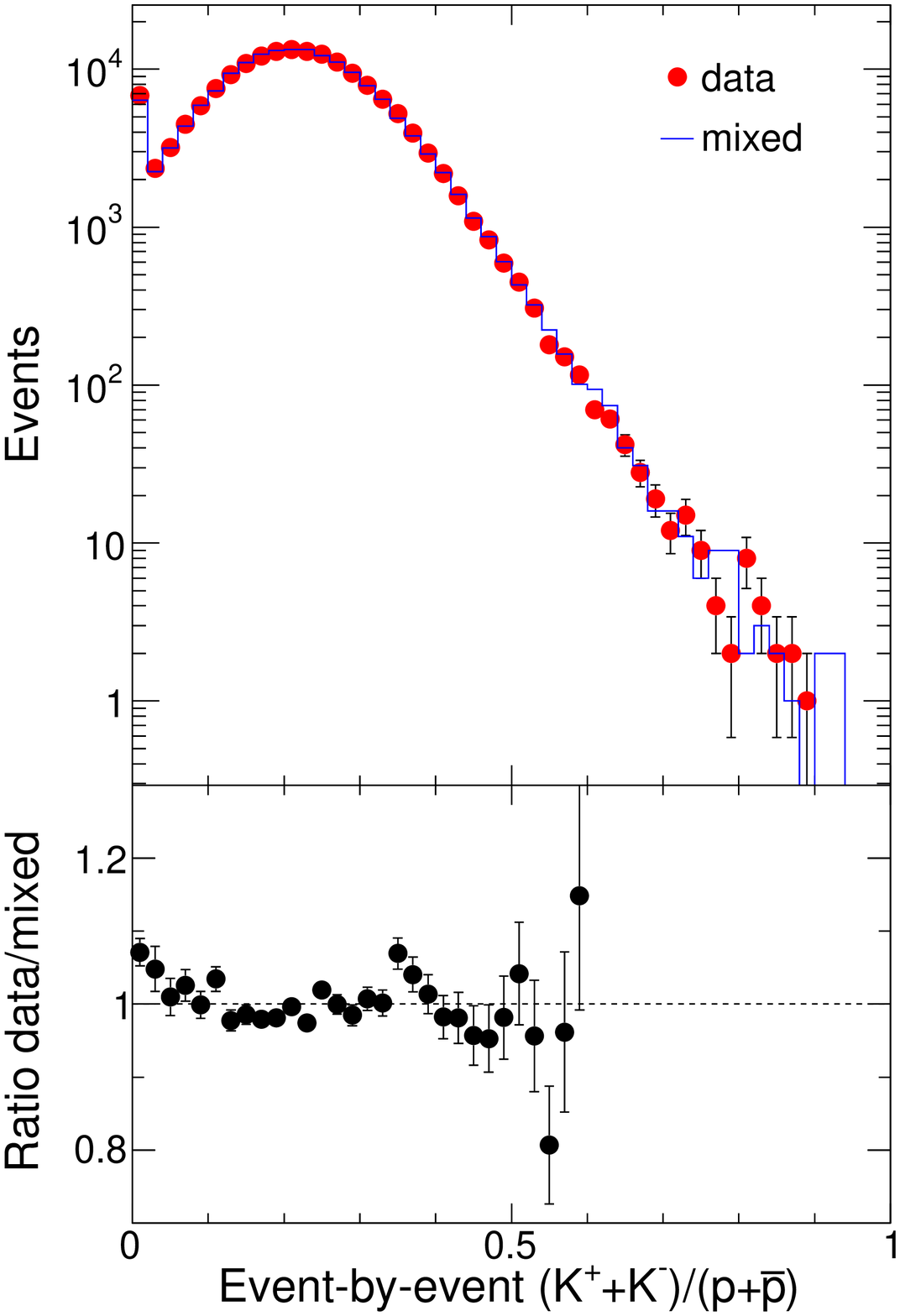}
\caption{(Color online) Event-by-event distribution of the \ktop\ ratio at \roots = 6.3~GeV for real data events compared to the mixed event reference. The lower panel shows the ratio data/mixed, where the concave shape indicates positive dynamical fluctuations.}
\label{fig:ebye_ktop_20}
\end{figure}

The fluctuations of the event-wise $\mathrm{K}/\mathrm{p}$ ratio are quantified using the scaled dispersion $\sigma := \sqrt{\mathrm{Var}(\mathrm{K}/\mathrm{p})} / \langle \mathrm{K}/\mathrm{p} \rangle$. The measured value in the data is denoted \sdata.
Figures~\ref{fig:ebye_ktop_160} and~\ref{fig:ebye_ktop_20} show the event-by-event distribution of the \ktop\ ratio for $\roots = 17.3$ and 6.3~GeV.
To quantify finite-number statistics and PID resolution effects, a reference sample of mixed events is constructed. This sample is made such as to preserve the original multiplicity distribution, with no two tracks in a mixed event taken from the same physical event. As described in detail in~\cite{:2008ca}, the mixed events contain no correlation due to physical processes, but effects from finite number statistics remain. The measured \dedx\ information is still attached to the individual particles, so that the likelihood method can be applied to the mixed events in the same way as to the original events. Thus, the effect of the \dedx\ resolution on the extracted particle ratios is reproduced by the mixed events.
The event-by-event distribution from mixed events is also displayed in Figs.~\ref{fig:ebye_ktop_160} and~\ref{fig:ebye_ktop_20}. Their scaled dispersion is denoted \smix.

The dynamical fluctuations can now be constructed as the quadratic difference~\cite{:2008ca,Afanasev:2000fu,Gazdzicki:1994vj}:
\begin{equation}
\sdyn := \mathrm{sign} \left( \sdata^2 - \smix^2 \right) \sqrt{\left| \sdata^2 - \smix^2 \right|}.
\end{equation}

At \roots = 17.3~GeV (Fig.~\ref{fig:ebye_ktop_160}) the data show a narrower distribution compared to mixed events, corresponding to a negative \sdyn. This is further visualized in the ratio between data and mixed event distributions shown in the lower panel of Fig.~\ref{fig:ebye_ktop_160} .
Conversely, at 6.3~GeV (Fig.~\ref{fig:ebye_ktop_20}) $\sdata > \smix$ and thus $\sdyn > 0$. Following from the less symmetrical shape of the event-by-event distribution here, this result is not so straight-forwardly inferred from the ratio plot.

The method described above has been successfully used and thoroughly tested in the analysis of \ktopi\ and $\mathrm{p}/\pi$ fluctuations~\cite{:2008ca}, and the same extensive quality checks were applied in the present analysis. For instance, outlying events with very small or high K/p ratios were found to contribute only modestly to the reported fluctuation signal: Consistent with~\cite{:2008ca}, the signal changes by less than 1\% when removing the high or low tails of the K/p distributions that correspond to 1\% of the events.
Special care was taken in the present analysis to check whether a correlation remains from the fit when extracting the $\mathrm{K}/\mathrm{p}$ ratio, as the \dedx\ based kaon-proton separation is smaller than for the kaon-pion case.
To exclude such an influence, events generated in the hadronic transport model UrQMD~\cite{Bleicher:1999xi,Bass:1998ca,Petersen:2008kb}
were used to study the effect of the \dedx\ resolution. The results of a direct model calculation are in agreement with those obtained using in addition a simulation of the NA49 TPC \dedx\ response and employing the fit procedure described above. The difference between the two methods amounts to 1.5\% at most and is taken into account in the systematic error.

The stability of the results was tested under variation of the track selection criteria, acceptance and event selection, as described in detail in~\cite{:2008ca}.
The changes due to these variations are also represented in the systematic errors. The results proved to be stable under small variations of the chosen acceptance. However, the acceptance tables provided in~\cite{acc_tab_edms} should be used for model comparisons.


\begin{figure}
\includegraphics[width=8cm]{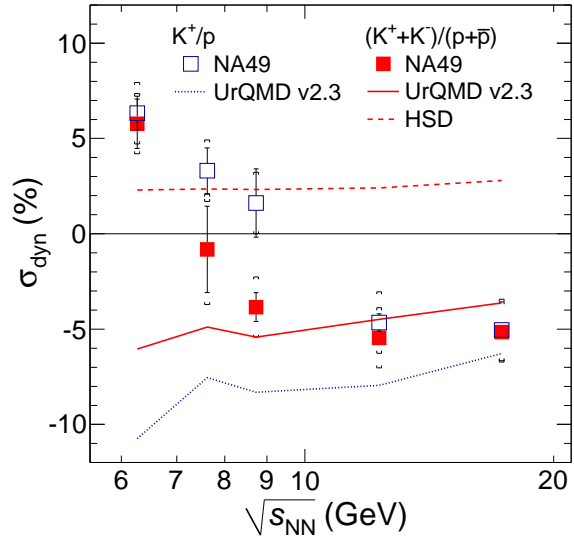}
\caption{(Color online) Energy dependence of the event-by-event dynamical fluctuations of the \ktop\ and the \ktopplus\ ratios. Symbols represent the NA49 measurements with statistical and systematic (braces) uncertainties. Calculations within the UrQMD and HSD transport models, processed through an NA49 acceptance filter are represented by lines. The statistical errors on the model results decrease from approximately 1.5\% at 6.3~GeV to 0.5\% at 17.3~GeV.}
\label{fig:edep_ktop}
\end{figure}

\sdyn\ has been evaluated for two ratios: \ktop\ and \ktopplus. The former probes correlations between all four involved hadron species. Only \ktopplus\ can be studied among the separate charge ratios, as the $\mathrm{K}^-$ and $\bar{\mathrm{p}}$ yields (see table~\ref{tbl:mult}) vanish at low energies. For the same reason, the two observables are expected to converge at low energies.
The excitation functions of \sdyn\ are shown in Fig.~\ref{fig:edep_ktop}.
For both cases, the dynamical ratio fluctuations change from positive (enhanced fluctuations compared to mixed events) at low SPS energies to negative values at the higher SPS energies. At high and low energies, \sdyn\ agrees for the two studied ratios, but disagrees at $\roots = 7.6$ and 8.7~GeV.
Recent preliminary results from the STAR collaboration~\cite{Tian:2010zzd} also find negative values and indicate a weak energy dependence between $\roots = 17.3$ and 200~GeV.

Calculations in the hadronic transport models UrQMD~\cite{Bleicher:1999xi,Bass:1998ca,Petersen:2008kb} and HSD~\cite{Ehehalt:1996uq,Konchakovski:2009at} are shown for comparison. The NA49 experimental acceptance was used in the model studies.
\sdyn\ shows only a weak energy dependence in both models. The striking change of sign and the strong energy dependence seen in the data are not reproduced.
In UrQMD, both charge combinations were evaluated and have a constant difference over the studied energy range, another feature in contrast to the data.


Unlike the previous results on \ktopi\ and \ptopi\ fluctuations, for which attempts of an explanation exist, the present data can not easily be understood. We recall that the \ptopi\ fluctuations were explained in hadronic models as a result of the proton-pion correlation due to resonance decay~\cite{:2008ca,Schuster:2009ak} and that the rise in \sdyn\ for \ktopi\ was suggested to be due to scaling properties of the observable \sdyn\ itself~\cite{Koch:2009dg} or might even be connected to the onset of deconfinement~\cite{Gorenstein:2003hk}.

As no known resonance feeds into positively charged kaons and protons, another source of correlation has to change at the energy where \sdyn\ switches sign, deviating from the transport model calculations. 
The change of sign is also incompatible with all scalings suggested in~\cite{Koch:2009dg} and thus indicative of a change in the underlying correlation physics.

In the baryon-strangeness correlation, a rapid change is expected at the deconfinement phase transition~\cite{Cheng:2008zh,Schmidt:2009qq} for which indications were found in the same energy region in rapid changes of several hadron production properties~\cite{Afanasiev:2002mx,:2007fe,Friese:CPOD09}.
The observed energy dependence qualitatively supports the scenario of a change in the baryon-strangeness correlation, but the exact contribution to \cbs\ of $\sdyn (\ktop)$ and $\sdyn (\ktopplus)$, respectively is still under discussion~\cite{Koch:pc}.


In summary, we present a first measurement of the dynamical fluctuations of the kaon-to-proton number ratio at the SPS energies.
Both \ktop\ and \ktopplus\ fluctuations show a non-trivial excitation function that is not reproduced in the hadronic models UrQMD and HSD, potentially pointing to a change in the baryon number-strangeness correlation at $\roots \approx 8$~GeV.
Although a connection between the kaon-to-proton ratio and \cbs\ seems reasonable, and the latter is suggested as a unique test for the basic degrees of freedom in the probed matter, the detailed connection between our measurement and \cbs\ and its interpretation require further theoretical studies.


This work was supported by
the US Department of Energy Grant DE-FG03-97ER41020/A000,
the Bundesministerium fur Bildung und Forschung, Germany,
German Research Foundation (grants GA 1480/2-1 and STO 205/4-1),
the Polish Ministry of Science and Higher Education (1 P03B 006 30, 1 P03B 127 30, 0297/B/H03/2007/33, N N202 078735, N N202 204638),
the Hungarian Scientific Research Foundation (T068506),
the Bulgarian National Science Fund (Ph-09/05),
the Croatian Ministry of Science, Education and Sport (Project 098-0982887-2878),
and Stichting FOM, the Netherlands.



\begin{thebibliography}{99}

\bibitem{Karsch:2003jg}
  F.~Karsch and E.~Laermann,
  arXiv:hep-lat/0305025.

\bibitem{Arnaldi:2008fw}
  R.~Arnaldi {\it et al.}  [NA60 Collaboration],
  Eur.\ Phys.\ J.\  C {\bf 61}, 711 (2009)
  [arXiv:0812.3053 [nucl-ex]].

\bibitem{Heinz:2000bk}
  U.~W.~Heinz and M.~Jacob,
  arXiv:nucl-th/0002042.

\bibitem{Gazdzicki:1998vd}
  M.~Gazdzicki and M.~I.~Gorenstein,
  Acta Phys.\ Polon.\  B {\bf 30}, 2705 (1999)
  [arXiv:hep-ph/9803462].

\bibitem{Afanasiev:2002mx}
  S.~V.~Afanasiev {\it et al.}  [The NA49 Collaboration],
  Phys.\ Rev.\  C {\bf 66}, 054902 (2002)
  [arXiv:nucl-ex/0205002].

\bibitem{:2007fe}
  C.~Alt {\it et al.}  [NA49 Collaboration],
  Phys.\ Rev.\  C {\bf 77}, 024903 (2008)
  [arXiv:0710.0118 [nucl-ex]].
  
\bibitem{Friese:CPOD09}
  V.~Friese,
  PoS {\bf CPOD09}, 005 (2009)

\bibitem{Gazdzicki:2004ef}
  M.~Gazdzicki {\it et al.}  [NA49 Collaboration],
  J.\ Phys.\ G {\bf 30}, S701 (2004)
  [arXiv:nucl-ex/0403023].

\bibitem{Fodor:2004nz}
  Z.~Fodor and S.~D.~Katz,
  JHEP {\bf 0404}, 050 (2004)
  [arXiv:hep-lat/0402006].

\bibitem{Ejiri:2003dc}
  S.~Ejiri, C.~R.~Allton, S.~J.~Hands, O.~Kaczmarek, F.~Karsch, E.~Laermann and C.~Schmidt,
  Prog.\ Theor.\ Phys.\ Suppl.\  {\bf 153}, 118 (2004)
  [arXiv:hep-lat/0312006].

\bibitem{Karsch:2007dp}
  F.~Karsch,
  PoS {\bf CPOD07}, 026 (2007)
  [arXiv:0711.0656 [hep-lat]].

\bibitem{Koch:2008ia}
  V.~Koch,
  arXiv:0810.2520 [nucl-th].

\bibitem{Zaranek:2001di}
  J.~Zaranek,
  Phys.\ Rev.\  C {\bf 66}, 024905 (2002)
  [arXiv:hep-ph/0111228].
  
\bibitem{Alt:2004ir}
  C.~Alt {\it et al.}  [NA49 Collaboration],
  Phys.\ Rev.\  C {\bf 70}, 064903 (2004)
  [arXiv:nucl-ex/0406013].
  
\bibitem{Koch:2005vg}
  V.~Koch, A.~Majumder and J.~Randrup,
  Phys.\ Rev.\ Lett.\  {\bf 95}, 182301 (2005)
  [arXiv:nucl-th/0505052].

\bibitem{Afanasev:1999iu}
  S.~Afanasiev {\it et al.}  [NA49 Collaboration],
  Nucl.\ Instrum.\ Meth.\  A {\bf 430}, 210 (1999).
  
\bibitem{:2008ca}
  C.~Alt {\it et al.}  [NA49 Collaboration],
  Phys.\ Rev.\  C {\bf 79}, 044910 (2009)
  [arXiv:0808.1237 [nucl-ex]].

\bibitem{acc_tab_edms}
https://edms.cern.ch/document/984431/1 

\bibitem{Afanasev:2000fu}
  S.~V.~Afanasiev {\it et al.}  [NA49 Collaboration],
  Phys.\ Rev.\ Lett.\  {\bf 86}, 1965 (2001)
  [arXiv:hep-ex/0009053].

\bibitem{Gazdzicki:1994vj}
  M.~Gazdzicki,
  Nucl.\ Instrum.\ Meth.\  A {\bf 345}, 148 (1994).

\bibitem{Bleicher:1999xi}
  M.~Bleicher {\it et al.},
  J.\ Phys.\ G {\bf 25}, 1859 (1999)
  [arXiv:hep-ph/9909407].

\bibitem{Bass:1998ca}
  S.~A.~Bass {\it et al.},
  Prog.\ Part.\ Nucl.\ Phys.\  {\bf 41}, 255 (1998)
  [arXiv:nucl-th/9803035].

\bibitem{Petersen:2008kb}
  H.~Petersen, M.~Bleicher, S.~A.~Bass and H.~Stocker,
  arXiv:0805.0567 [hep-ph].

\bibitem{Tian:2010zzd}
  J.~Tian  [STAR Collaboration],
  J.\ Phys.\ G {\bf 37}, 094044 (2010).

\bibitem{Ehehalt:1996uq}
  W.~Ehehalt, W.~Cassing,
  Nucl.\ Phys.\  {\bf A602}, 449-486 (1996).
  
\bibitem{Konchakovski:2009at}
  V.~P.~Konchakovski, M.~Hauer, M.~I.~Gorenstein, E.~L.~Bratkovskaya,
  J.\ Phys.\ G {\bf 36}, 125106 (2009).
  [arXiv:0906.3229 [nucl-th]].

\bibitem{Schuster:2009ak}
  T.~Anticic {\it et al.}  [NA49 Collaboration],
  PoS {\bf CPOD2009}, 029 (2009)
  [arXiv:0910.0558 [nucl-ex]].

\bibitem{Koch:2009dg}
  V.~Koch and T.~Schuster,
  Phys.\ Rev.\  C {\bf 81}, 034910 (2010)
  [arXiv:0911.1160 [nucl-th]].

\bibitem{Gorenstein:2003hk}
  M.~I.~Gorenstein, M.~Gazdzicki and O.~S.~Zozulya,
  Phys.\ Lett.\  B {\bf 585}, 237 (2004)
  [arXiv:hep-ph/0309142].

\bibitem{Cheng:2008zh}
  M.~Cheng {\it et al.},
  Phys.\ Rev.\  D {\bf 79}, 074505 (2009)
  [arXiv:0811.1006 [hep-lat]].

\bibitem{Schmidt:2009qq}
  C.~Schmidt,
  PoS {\bf CPOD2009}, 024 (2009)
  [arXiv:0910.4321 [hep-lat]].

\bibitem{Koch:pc}
  V.~Koch, private communication.

\end{thebibliography}
\end{document}